\documentclass[sigconf,9pt]{acmart}
\usepackage[english]{babel}
\usepackage{blindtext}
\usepackage{caption}
\usepackage{subcaption}
\usepackage{color}
\usepackage{tabularx}
\usepackage{multirow}
\usepackage{multicol}
\usepackage{graphicx}
\usepackage{textgreek}
\usepackage{tabularray}
\usepackage{float}
\usepackage{xcolor}
\usepackage{todonotes}
\usepackage{array} % To use the 'p' column type
\usepackage{makecell} % To allow for line breaks within cells
\usepackage{booktabs} % For better table lines
\usepackage{geometry}
\usepackage{microtype}
\usepackage{booktabs} 
\usepackage{placeins} 
\setcopyright{acmlicensed}

\copyrightyear{2026}
\acmYear{2026}
\setcopyright{cc}
\setcctype{by}
\acmConference[SenSys '26]{ACM/IEEE International Conference on Embedded Artificial Intelligence and Sensing Systems}{May 11--14, 2026}{Saint Malo, France}
\acmBooktitle{ACM/IEEE International Conference on Embedded Artificial Intelligence and Sensing Systems (SenSys '26), May 11--14, 2026, Saint Malo, France}
\acmDOI{10.1145/3774906.3802782}
\acmISBN{979-8-4007-2309-4/2026/05}

\begin{document}

\title{Short Paper: WearBCI Dataset: Understanding and Benchmarking Real-World Wearable\\ Brain-Computer Interfaces Signals}

% -------------------------
% Compact author block
% -------------------------
\author{Haoxian Liu, Hengle Jiang, Lanxuan Hong, Xiaomin Ouyang}
\authornote{Corresponding author}
\affiliation{
  \institution{Hong Kong University of Science and Technology}
  \country{Hong Kong SAR, China}
}
\email{hliueu@connect.ust.hk, hjiangbg@connect.ust.hk, lhongae@connect.ust.hk, xmouyang@cse.ust.hk}

\begin{abstract}
Brain-computer interfaces (BCIs) have opened new platforms for human-computer interaction, medical diagnostics, and neurorehabilitation. Wearable BCI systems, which typically employ non-invasive electrodes for portable monitoring, hold great promise for real-world applications, but also face significant challenges of signal quality degradation caused by motion artifacts and environmental interferences. Most existing wearable BCI datasets are collected under stationary or controlled lab settings, limiting their utility for evaluating performance under body movement. To bridge this gap, we introduce WearBCI, the first dataset that comprehensively evaluates wearable BCI signals under different motion dynamics with synchronized multimodal recordings (EEG, IMU, and egocentric video), and systematic benchmark evaluations for studying impacts of motion artifact. Specifically, we collect data from 36 participants across different motion dynamics, including body movements, walking, and navigation. This dataset includes synchronized electroencephalography (EEG), inertial measurement unit (IMU) data, and egocentric video recordings. We analyze the collected wearable EEG signals to understand the impact of motion artifacts across different conditions, and benchmark representative EEG signal enhancement techniques on our dataset. Furthermore, we explore two new case studies: cross-modal EEG signal enhancement and multi-dimension human behavior understanding. These findings offer valuable insights into real-world wearable BCI deployment and new applications\footnote{The dataset and code are available at: https://github.com/HKUST-MINSys-Lab/WearBCI-Dataset.}.
\end{abstract}




\maketitle

\section{Introduction}
% EEG has been widely applied for diagnosing neurological diseases \cite{cassani2018systematic, babiloni2021measures, miladinovic2020evaluation}, and neurofeedback training\cite{enriquez2017eeg}. 
EEG-based brain-computer interfaces (BCIs) have enabled applications in human-computer interaction \cite{karikari2023review, alimardani2020passive}, neurorehabilitation \cite{enriquez2017eeg, orban2022review, lazarou2018eeg}, and medical diagnostics \cite{vidyaratne2017real, van2014functional}. However, conventional BCI systems are expensive and require complex electrode setup, making them impractical for long-term use in natural daily environments.
To overcome these limitations, wearable BCIs have emerged as a promising technology by providing non-invasive and portable solutions for applications like attention regulation \cite{huang2024real, belo2021eeg}, sleep monitoring \cite{ferster2022benchmarking, aboalayon2016sleep, koushik2018real}, and stroke rehabilitation \cite{qin2019econhand}.

Despite their promise, wearable BCIs still face major challenges for real-world deployment. Wearable BCI systems are highly vulnerable to motion-induced artifacts: facial and scalp muscles generate strong EMG signals that overlap with neural activity, and body movements cause subtle electrode shifts, disrupting skin contact and increasing impedance, introducing voltage jumps that degrade signal quality \cite{chi2010dry, mathewson2017high}.
However, existing wearable BCI datasets are largely collected under stationary, controlled conditions (e.g., sitting or resting), overlooking movement and environmental variability. This gap limits the development and evaluation of wearable BCI systems for everyday scenarios.

In this paper, we present WearBCI, the first dataset that comprehensively evaluates wearable BCI signals under different motion dynamics with synchronized multimodal recordings (EEG, IMU, and egocentric video), and systematic benchmark evaluations for studying impacts of motion artifact. To systematically assess the impact of motion on EEG signal quality, we designed experiments at three levels of motion intensity: body movements, walking, and navigation. The dataset includes recordings from 36 participants, consisting of approximately 17 hours of EEG, 10 hours of IMU, and 1 hour of egocentric video data. The video recordings are collected only during the navigation session, while EEG and IMU are available across all motion sessions.

We then analyze the impact of motion conditions on EEG signal quality using time-domain features, power spectral density (PSD), and topographic maps across brain regions. Results show that EEG signals remain stable under mild movement, but stronger acceleration introduces larger time-domain fluctuations and broadband spectral elevation. Different actions also affect distinct scalp regions, highlighting the need for motion- and region-aware processing.
To benchmark existing EEG enhancement methods under motion conditions, we apply six representative approaches on our dataset: three traditional methods, including Independent Component Analysis (ICA), Empirical Mode Decomposition (EMD), and Artifact Subspace Reconstruction (ASR), and three deep learning models (GAN, Transformer, and Diffusion-based). Results show that classical methods degrade as motion intensity increases, while deep models better restore spectra but risk suppressing neural signals. Moreover, analysis across different navigation scenes shows that abrupt behaviors, such as obstacle avoidance, introduce more severe distortions than smoother actions.

To further showcase new applications enabled by our dataset, we present two case studies. The first case study explores how additional sensor modalities, such as IMU signals, can assist in enhancing EEG quality under motion conditions. The second case study motivates the practical value of the WearBCI dataset by demonstrating application scenarios like comprehensive human behavior understanding enabled by multimodal signals. These findings highlight the potential of our dataset for addressing challenges in real-world deployment with various motion artifacts and enable new applications of wearable BCI systems.

In summary, we make the following key contributions:
\begin{itemize}
    \item We introduce the first multimodal wearable BCI dataset with synchronized recordings (EEG, IMU, and egocentric video) from 36 subjects across different motion dynamics, covering body movements, walking, and navigation.
    \item We benchmark the collected wearable EEG data by evaluating the impact of various artifacts across sessions and assessing representative EEG enhancement methods, providing insights for motion-aware wearable BCI applications.
    \item We conduct two case studies including cross-modal EEG signal enhancement using complementary sensor modalities such as IMU, and multi-dimensional behavior analysis using multimodal data and large language models.
\end{itemize}

\section{Related Work}
\textbf{EEG-Based BCI Systems.}
EEG-based BCIs provide signals with high temporal resolution, valuable for identifying biomarkers of neurodegenerative diseases like Alzheimer's \cite{cassani2018systematic,babiloni2021measures,ouyang2024admarker} and Parkinson's \cite{yang2019m,miladinovic2020evaluation}, and for neurofeedback in rehabilitation \cite{orban2022review,lazarou2018eeg}. However, traditional systems often require complex, invasive setup, limiting their practicality in daily environments. Wearable BCIs overcome these barriers with portable solutions for attention tracking \cite{snoek2025wearables,huang2024real,belo2021eeg}, sleep monitoring \cite{ferster2022benchmarking,aboalayon2016sleep,koushik2018real}, and motor rehabilitation \cite{sayegh2017wearable,arpaia2020wearable}, despite remaining prone to motion artifacts and bioelectrical interferences \cite{kumaravel2021efficient,kim2015real}.

\noindent \textbf{Wearable BCI Datasets.}
A growing number of wearable BCI datasets have been collected for diverse applications. BCIC IV 2b \cite{leeb2008bci} and MILimbEEG \cite{asanza2023milimbeeg} target motor control and rehabilitation, while AMIGOS \cite{miranda2018amigos} and Emognition \cite{saganowski2022emognition} combine EEG with GSR for emotion recognition. For multimodal datasets, MOCAS \cite{jo2024mocas} captures EEG, PPG, GSR, and video data during anomaly detection tasks, and MPDB \cite{tao2024multimodal} collects EEG, ECG, EMG, GSR, and eye movement data in a driving simulator for behavior classification. However, both datasets focus on task-specific settings without systematic variation of body motion. Overall, most existing datasets are collected under stationary or controlled conditions, and none systematically study how different motion dynamics degrade wearable EEG with synchronized IMU and egocentric video. WearBCI addresses this gap by providing the first multimodal dataset designed specifically for benchmarking motion artifact analysis and EEG enhancement under different motion dynamics.

% In this section, we should first analyze the differe nces between wearable and clinical BCI devices/ signals from the principles, then analyze/compare data collected by the two types of devices, fina lly summarize the key challenges of interpreting wearable BCI signals.
% Need a figure to compare wearable and clinical BCI devices and signals

\section{Background and challenges}

\begin{figure}
    \centering
    \setlength{\abovecaptionskip}{0.cm}
    \setlength{\belowcaptionskip}{0.cm}
    \includegraphics[width=0.98\linewidth]{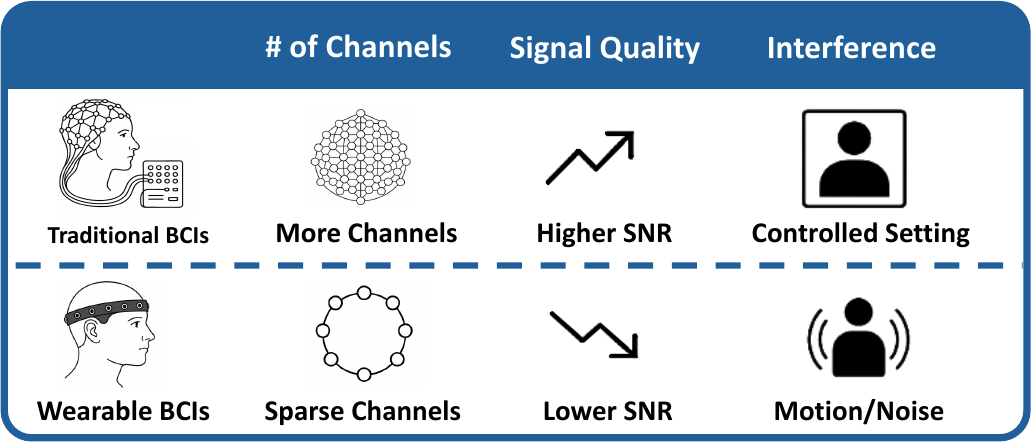}
    \caption{Comparison of traditional and wearable BCIs. 
    }
    \label{fig:BCI_devices_compare}
\end{figure}

In this section, we compare traditional and wearable BCI systems, highlighting key challenges of using wearable EEG in real-world environments (Figure~\ref{fig:BCI_devices_compare}).

\noindent \textbf{Sparse Electrode Channels. }
Wearable BCIs typically use 1-8 dry electrodes on frontal or occipital areas \cite{zhang2023recent}, allowing quick setup but limiting spatial resolution and brain coverage. In contrast, traditional systems deploy 21-256 wet electrodes \cite{rao2013brain} for full-head monitoring, but require 20-60 minutes for setup and adjustment \cite{zander2011dry}, making them less suitable for real-world deployment.

\noindent \textbf{Lower SNR.}
Signal quality is another major challenge. Traditional BCIs use wet electrodes with low impedance (<20 k$\Omega$), yielding high-fidelity signals \cite{fiedler2015novel}, whereas wearable systems use dry electrodes at the cost of increased impedance and reduced SNR due to greater susceptibility to noise and contact instability \cite{ratti2017comparison}.

\noindent \textbf{Susceptibility to Interference.}
Wearable BCIs are highly vulnerable to motion-induced artifacts. Facial and scalp muscles generate strong EMG signals that overlap with neural activity, while body movements cause subtle electrode shifts that disrupt skin contact, increase impedance, and introduce voltage jumps that degrade signal quality. In contrast, traditional BCIs operate in controlled environments with stable wet-electrode contact, making them far less susceptible to such interference.
\section{WearBCI Dataset Collection}

\subsection{Experiment Setup}

% \textbf{Devices and collected data.}
% EEG data were collected using the OpenBCI Cyton board \cite{openbciCytonBiosensing} with eight dry electrodes positioned at Fp1, Fp2, P7, P8, T3, T4, O1, and O2 (reference and ground at A1 and A2), following the 10-20 system \cite{niedermeyer2005electroencephalography}. The overall setup is illustrated in Fig.\ref{fig:experiment_setup}. Electrode impedance was maintained below 50 k$\Omega$. As dry electrodes under motion conditions are prone to contact changes that increase impedance, we check the impedance before each session type and reposition the electrodes when it exceeded 50 k$\Omega$. IMUs were placed on the head, wrists, and ankles for motion artifact analysis, and egocentric video was recorded during navigation. All data streams were synchronized using the Network Time Protocol (NTP), with EEG recorded at 250 Hz, IMU at 100 Hz, and egocentric video at 30 FPS, each accompanied by Unix timestamps and aligned via timestamp-based interpolation. Preprocessing included a 1 Hz high-pass filter for slow drift removal, a 45 Hz low-pass filter, and ICA-based artifact removal. Due to the limited 8 channels, ICA was restricted to 8 components, so only the most prominent 1-2 artifact components were removed to preserve signal fidelity. Table~\ref{tab:dataset_stats} summarizes the dataset statistics.

\begin{figure}
    \centering
    \setlength{\abovecaptionskip}{0.cm}
    \setlength{\belowcaptionskip}{0.cm}

    % ---- Row 1: (a) and (b) ----
    \begin{subfigure}[b]{0.7\linewidth}
        \centering
            \setlength{\abovecaptionskip}{0.cm}
    \setlength{\belowcaptionskip}{0.cm}

        \includegraphics[width=\linewidth]{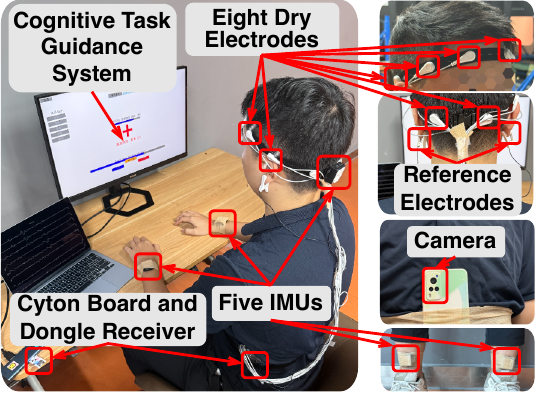}
        \caption{Experimental setup.}
        \label{fig:experiment_setup}
    \end{subfigure}
    % \hfill
    \begin{subfigure}[b]{0.23\linewidth}
        \centering
            \setlength{\abovecaptionskip}{0.cm}
    \setlength{\belowcaptionskip}{0.cm}

        \includegraphics[width=\linewidth]{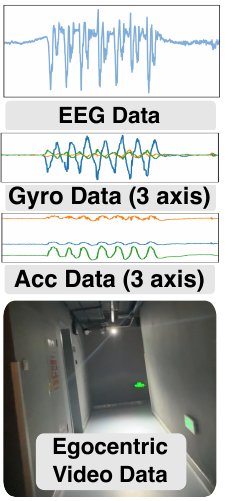}
        \caption{Data.}
        \label{fig:data}
    \end{subfigure}

    % ---- Row 2: (c) ----
    % \vspace{0.1cm}
    \begin{subfigure}[b]{0.93\linewidth}
        \centering
        \includegraphics[width=\linewidth]{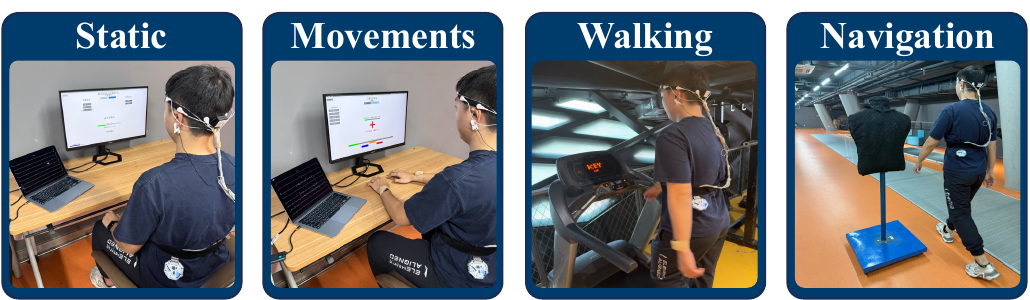}
        \caption{Illustration of different experiment sessions.}
        \label{fig:session}
    \end{subfigure}

    \caption{Overview of the WearBCI experimental setting and recorded multimodal data.}
    \label{fig:experiment_setting_merged}
\end{figure}

\noindent \textbf{Devices and collected data.}
EEG data were collected using the OpenBCI Cyton board \cite{openbciCytonBiosensing} with eight dry electrodes positioned at Fp1, Fp2, P7, P8, T3, T4, O1, and O2 (reference and ground at A1 and A2), following the 10-20 system \cite{niedermeyer2005electroencephalography}. The overall setup is illustrated in Fig.\ref{fig:experiment_setup}. Electrode impedance was maintained below 50 k$\Omega$. As dry electrodes under motion conditions are prone to contact changes that increase impedance, we check the impedance before each session type and reposition the electrodes when it exceeded 50 k$\Omega$. IMUs were placed on the head, wrists, and ankles for motion artifact analysis, and egocentric video was recorded during navigation. All data streams were synchronized using the Network Time Protocol (NTP) to establish a common time reference, with EEG recorded at 250 Hz, IMU at 100 Hz, and egocentric video at 30 FPS, each accompanied by Unix timestamps. Post-processing aligns all streams through timestamp-based interpolation. Preprocessing included a 1 Hz high-pass filter for slow drift removal, a 45 Hz low-pass filter for high-frequency noise, and ICA-based artifact removal. Due to the limited 8 channels, ICA was restricted to 8 components, so only the most prominent 1-2 artifact components were removed to preserve signal fidelity. Due to technical issues, data from 3 participants in the walking session and 4 participants in the navigation session were excluded from analysis. Occasional packet loss resulted in missing IMU samples across several recordings. Table~\ref{tab:dataset_stats} summarizes the dataset statistics.

\noindent \textbf{Subject recruitment.}
36 healthy adults (ages 18–26, balanced gender ratio, right-handed) were recruited\footnote{All the data collection was approved by the Institutional Review Board (IRB) of the authors' institution}.
Individuals with neurological, psychological, or cardiovascular disorders, or taking medications affecting brain activity, were excluded. Participants avoided alcohol, excessive caffeine, and sleep deprivation within 24 hours prior, and all provided informed consent.

\subsection{Experiment Protocol}
This study evaluates wearable BCIs under different motion dynamics. The three motion sessions were designed to cover discrete body actions, continuous locomotion, and a naturalistic navigation scenario combining movement with environmental interaction, providing a structured setting for examining how motion complexity affects EEG signal quality. Paired EEG and IMU were recorded across all non-static sessions, with egocentric video additionally collected during navigation. To reduce cognitive variability, participants were instructed to remain relaxed during the first three sessions; in the navigation task, cognitive load was kept minimal despite simple spatial decisions (e.g., turning or stopping).

\noindent \textbf{Static baseline. }
EEG was recorded while participants sat still to establish a physiological reference. Participants performed eyes-open and eyes-closed tasks lasting two minutes, each repeated three times.

\noindent \textbf{Body movements.}
This session examined the influence of small-scale body and facial movements on EEG. Prior wearable EEG literature \cite{amin2024motion, seok2021motion} has identified a broad set of motion types that introduce artifacts, covering head motions, eye actions, and upper-body movements. From this set, we selected five representative actions: head shaking, head nodding, eye movement, arm stretching, and typing, allowing isolated examination of how different body parts affect distinct scalp regions. Each action followed 10-second cycles of baseline, motion, and post-motion, repeated three times.

\noindent \textbf{Walking.}
This session evaluated the effect of continuous walking on EEG. Participants walked at speeds of 2, 3, and 4 km/h, representing slow, moderate, and brisk pedestrian movement to examine how gait intensity affects signal degradation, with two minutes per trial and two-minute rest periods between trials. Baseline recordings were collected under standing conditions as reference.

\noindent \textbf{Natural navigation.}
This session recorded neural activity during navigation tasks in an indoor setting. Participants followed a predefined route consisting of six sequential tasks: opening a door, greeting an experimenter, avoiding an obstacle, walking along a corridor, reading a wall poster, and knocking on a door. This design ensures that all participants encounter the same set of action types and environmental contexts, enabling systematic comparison of EEG responses across scenes. Walking speed and pause duration were unconstrained to preserve natural movement, with 10-second standing baselines at the start and end. The entire sequence took approximately 1.5 minutes.

\begin{table}
\setlength{\abovecaptionskip}{0.cm}
\setlength{\belowcaptionskip}{0.cm}
\centering
\resizebox{\columnwidth}{!}{
\begin{tabular}{l|c}
\toprule
\textbf{Subjects} & 36 healthy adults (18 - 26 yrs), balanced gender ratio \\
\midrule
\textbf{Total Duration} & EEG: 16.4 h,\ IMU: 9.2 h,\ Video: 0.9 h \\
\midrule
\textbf{Modalities} & 8-ch EEG (250 Hz), 5 IMUs (100 Hz), egocentric video (30 FPS) \\
\midrule
\textbf{Sessions} & Static, Body Movements (5 types), Walking (3 speeds), Navigation \\
\bottomrule
\end{tabular}
}
\caption{Summary of the WearBCI Dataset.}
\label{tab:dataset_stats}
\end{table}
\section{Evaluation}

\subsection{Understanding the WearBCI Dataset}
\label{sec:understand}

\subsubsection{Association between IMU and EEG signal.}
\begin{figure}
    \centering
    \setlength{\abovecaptionskip}{0.cm}
    \setlength{\belowcaptionskip}{0.cm}

    \begin{subfigure}{0.49\linewidth}
        \centering
        \includegraphics[width=\linewidth]{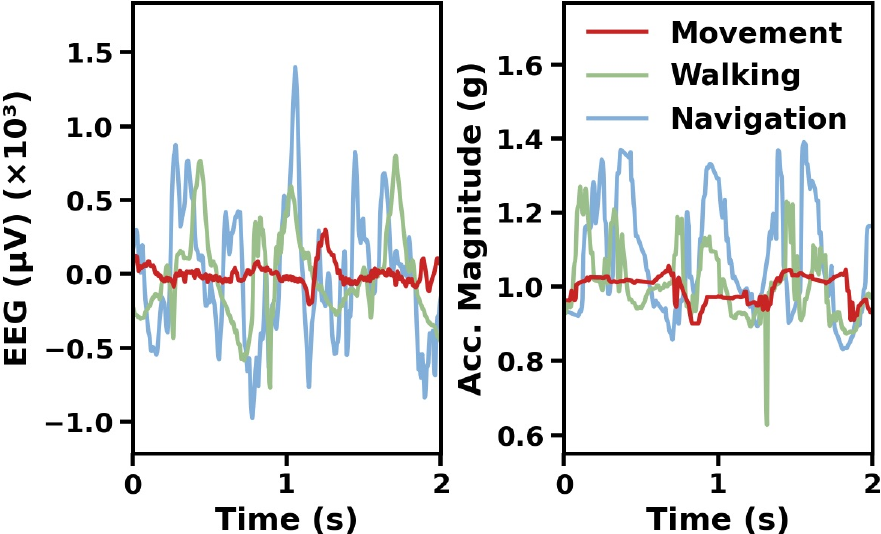}
        \caption{EEG and IMU waveforms.}
        \label{fig:imu-eeg-time}
    \end{subfigure}
    \hfill
    \begin{subfigure}{0.50\linewidth}
        \centering
        \includegraphics[width=\linewidth]{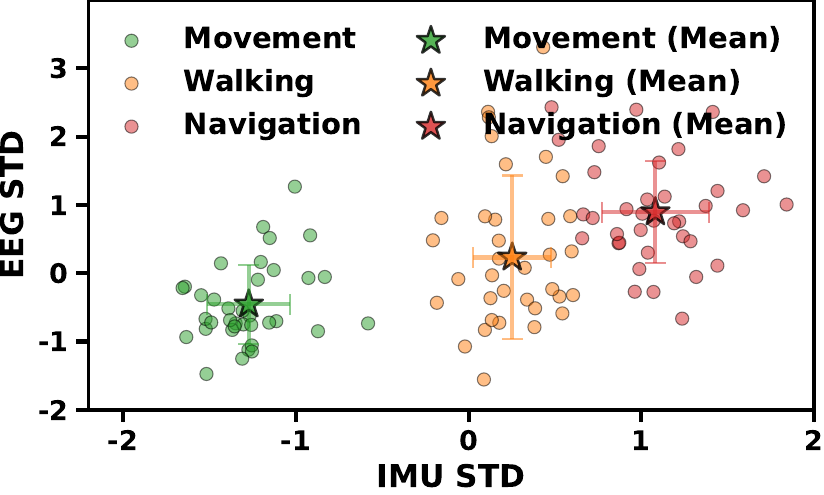}
        \caption{IMU-EEG correlation.}
        \label{fig:eeg-imu-correlation}
    \end{subfigure}

    \caption{Time-domain waveforms and motion-EEG correlation across different motion levels.}
    \label{fig:time-domain}
\end{figure}

Figure~\ref{fig:imu-eeg-time} shows raw EEG and IMU signals across motion conditions. As motion intensity increases, both signals exhibit larger amplitudes and variability: EEG displays stronger fluctuations due to electrode shifts and scalp muscle activation, walking introduces rhythmic gait oscillations, and navigation induces irregular bursts. Figure~\ref{fig:eeg-imu-correlation} quantifies this by computing STD of IMU and EEG signals across all 36 participants, revealing a clear association between IMU and EEG variability. Physically, increased acceleration leads to transient electrode-contact changes and scalp muscle activity, supporting the use of IMU-derived features as proxies for motion-induced EEG degradation and motivating IMU-aware denoising strategies.

\subsubsection{Impact on spectral characteristics.}
\begin{figure}
    \centering
    \setlength{\abovecaptionskip}{0.cm}
    \setlength{\belowcaptionskip}{0.cm}

    \begin{subfigure}{0.32\linewidth}
        \centering
        \includegraphics[width=\linewidth]{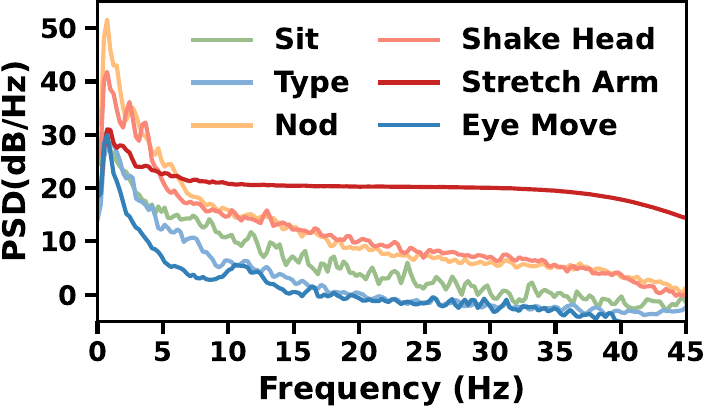}
        \caption{Body movements.}
        \label{fig:psd-daily-action}
    \end{subfigure}
    \hfill
    \begin{subfigure}{0.32\linewidth}
        \centering
        \includegraphics[width=\linewidth]{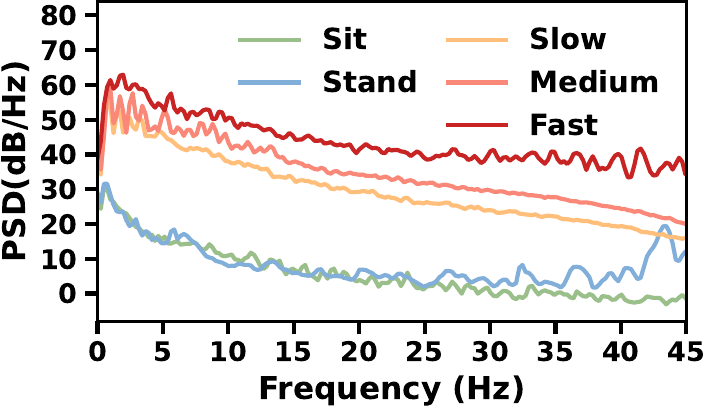}
        \caption{Walking speeds.}
        \label{fig:psd-walking}
    \end{subfigure}
    \hfill
    \begin{subfigure}{0.32\linewidth}
        \centering
        \includegraphics[width=\linewidth]{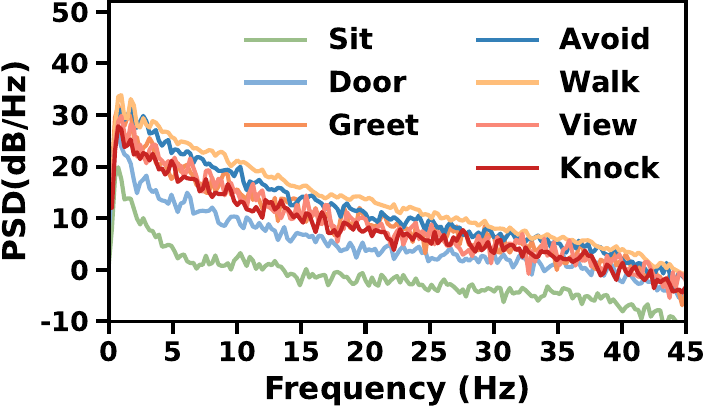}
        \caption{Navigation.}
        \label{fig:psd-navigation}
    \end{subfigure}

    \caption{PSD of EEG signals in different settings.}
    \label{fig:PSD}
\end{figure}

We compute PSD across motion dynamics to assess spectral distortions. Resting-state EEG shows a clear alpha peak (8-13 Hz) during eyes-closed, consistent with the BCIC IV 2a dataset~\cite{brunner2008bci}, supporting the physiological reliability of WearBCI under static conditions. As shown in Figure \ref{fig:PSD}, distortion increases with motion: light activities introduce mild broadband elevation, while larger head movements lead to stronger increases. During walking, PSD elevation grows progressively with speed. Under navigation, distortion varies by scene—tasks like walking or knocking cause greater broadband elevation, while standing or brief avoidance actions show milder changes.

\subsubsection{Impact on different brain regions.}
\label{sec:region}
\begin{figure}
    \centering
    \setlength{\abovecaptionskip}{0.cm}
    \setlength{\belowcaptionskip}{0.cm}
    \includegraphics[width=0.8\linewidth]{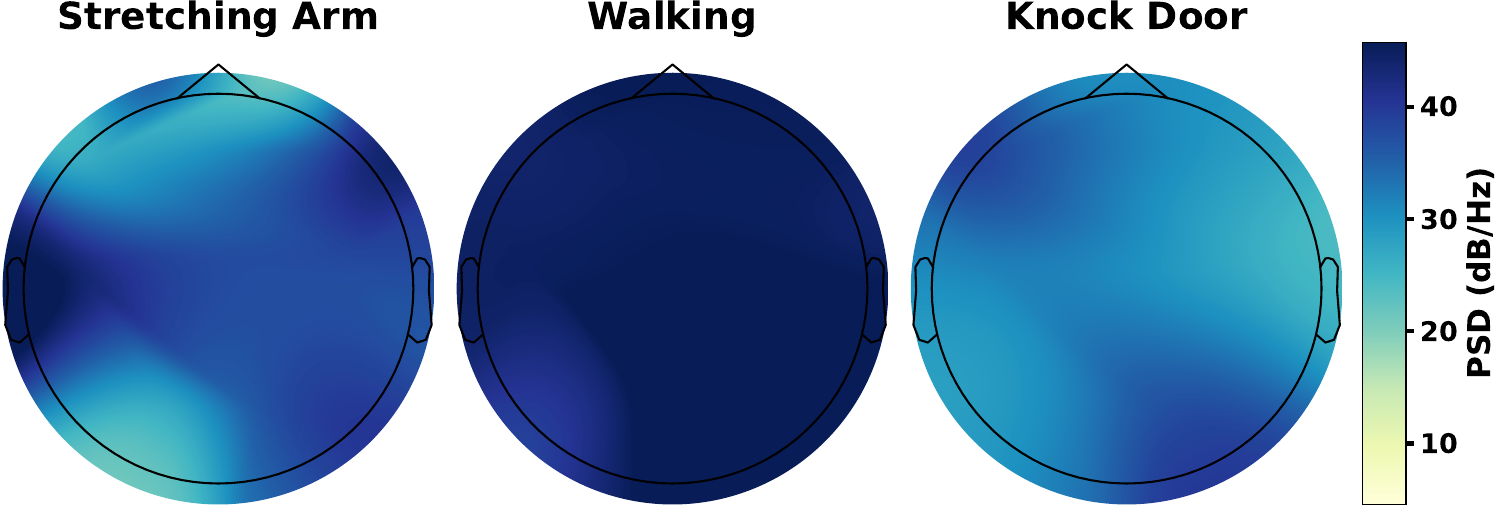}
    \caption{Topographic maps of different brain regions. Colors indicate PSD values, blue is higher.}
    \label{fig:topomap}
\end{figure}

We analyze how motion artifacts influence the spatial distribution of alpha-band power by computing topographic maps(Figure~\ref{fig:topomap}). The results show distinct regional effects depending on the type of movement. Arm stretching increases power in parietal and temporal regions, knocking affects frontal and occipital areas, and walking causes widespread elevation across all regions. These findings suggest that contamination patterns depend on both motion intensity and action type.

\subsection{Benchmarking EEG Enhancement Approaches}

We then evaluate the performance of several EEG denoising algorithms on our dataset.

\subsubsection{Methodology} We evaluate the following EEG enhancement algorithms.

\noindent \textbf{ICA:} A blind source separation technique that decomposes EEG into statistically independent components. Artifacts are removed by discarding components identified via manual inspection.

\noindent \textbf{EMD:} A signal decomposition method that extracts intrinsic mode functions (IMFs). Components dominated by noise or high entropy are selectively suppressed to recover cleaner waveforms.

\noindent \textbf{ASR:} A statistical technique that identifies and suppresses artifacts by reconstructing the signal within a low-variance subspace, based on deviations from clean calibration segments.

\noindent \textbf{GCTNet (GAN-based) \cite{yin2023gan}: }A hybrid network combining convolutional and transformer branches within a GAN framework. The generator restores EEG signals, while the discriminator ensures global signal realism through adversarial training.

\noindent \textbf{EEGDNet (Transformer-based) \cite{pu2022eegdnet}: }An attention-based network that models both local and global dependencies via transformer layers, enabling fine-grained removal of motion noise.

\noindent \textbf{EEGDfus (Diffusion-based) \cite{huang2024eegdfus}: }A conditional diffusion model that iteratively refines noisy EEG segments toward clean targets, leveraging multi-scale attention across adjacent windows for fine-grained temporal recovery.

\subsubsection{Evaluation Metrics.} We evaluate denoising performance using four metrics covering spectral fidelity, physiological consistency, artifact suppression, and signal interpretability.

\noindent \textbf{Weighted PSD Distance ($D_{\text{PSD}}$):}
Following \cite{zhang2021eegdenoisenet}, this metric measures $L_2$ spectral deviation of the denoised signal against static eyes-closed reference over 1-45 Hz. We use double weighting on alpha (8-13 Hz) and beta (13-30 Hz) bands to emphasize key neural rhythms. Lower is better.

\noindent \textbf{Alpha Peak Consistency ($D_{\alpha}$):}
This metric measures the deviation of the alpha peak amplitude of the denoised signal against the static eyes-closed baseline \cite{steyrl2018rlaf}, assessing whether key physiological oscillations are preserved after denoising. Lower values indicate better retention of spectral structure.

\noindent \textbf{EMG Index:}
Ratio of high-frequency (30-45 Hz) to low-frequency (8-30 Hz) power in the denoised signal, reflecting residual muscle artifact contamination \cite{modarres2022_hflf_emg}. Lower values imply cleaner signals.

\noindent \textbf{Brain IC Ratio:}
Proportion of components after ICA decomposition labeled as ``Brain'' by ICLabel \cite{pion2019iclabel}, indicating how well neural sources are preserved after denoising. Higher is better. When denoising effectively suppresses artifacts while retaining neural signals, ICA can identify a higher proportion of brain-origin components. This metric thus captures neural preservation.

\begin{table*}
\centering
\setlength{\tabcolsep}{3pt}
\renewcommand{\arraystretch}{1.05}
\resizebox{\textwidth}{!}{
\begin{tabular}{l|cccc|cccc|cccc}
\toprule
\multirow{2}{*}{\textbf{Method}}
& \multicolumn{4}{c|}{\textbf{Body Movements}}
& \multicolumn{4}{c|}{\textbf{Walking}}
& \multicolumn{4}{c}{\textbf{Navigation}} \\
& $D_{\mathrm{PSD}}$ & $D_\alpha$ & EMG Index & Brain
& $D_{\mathrm{PSD}}$ & $D_\alpha$ & EMG Index & Brain
& $D_{\mathrm{PSD}}$ & $D_\alpha$ & EMG Index & Brain \\
\midrule
ICA     & $16.1\pm 15.5$ & \underline{$2.1\pm 0.4$} & $1.3\pm 1.8$ & $\mathbf{0.47\pm 0.19}$
        & $31.5\pm 16.2$ & $\mathbf{1.6\pm 0.4}$ & \underline{$1.1\pm 9.3$} & $0.30\pm 0.17$
        & $50.6\pm 34.9$ & $\mathbf{1.7\pm 0.6}$ & $1.7\pm 0.4$ & \underline{$0.50\pm 0.35$} \\
EMD     & $25.0\pm 19.4$ & $\mathbf{1.7\pm 0.3}$ & $\mathbf{0.2\pm 0.1}$ & \underline{$0.38\pm 0.15$}
        & $96.4\pm 40.3$  & $3.4\pm 0.4$ & $\mathbf{0.3\pm 0.2}$ & \underline{$0.38\pm 0.18$}
        & $213.9\pm 167.9$& $1.9\pm 0.4$ & $\mathbf{0.5\pm 0.1}$ & $0.41\pm 0.12$ \\
ASR     & $12.8\pm 12.6$ & $2.1\pm 0.5$ & $1.2\pm 0.7$ & $0.28\pm 0.21$
        & $23.6\pm 7.5$   & $2.2\pm 0.7$ & $2.3\pm 2.9$ & $\mathbf{0.45\pm 0.23}$
        & $145.0\pm 155.1$& \underline{$1.8\pm 0.3$} & $1.6\pm 0.2$ & $\mathbf{0.66\pm 0.16}$ \\
GCTNet  & $13.7\pm 11.2$ & $2.3\pm 0.3$ & $2.0\pm 0.4$ & $0.20\pm 0.10$
        & $16.4\pm 62.8$  & \underline{$2.1\pm 0.4$} & $2.0\pm 0.4$ & $0.33\pm 0.20$
        & \underline{$9.0\pm 3.5$}    & $2.4\pm 0.2$ & $1.9\pm 0.1$ & $0.19\pm 0.22$ \\
EEGDnet & \underline{$9.5\pm 3.2$}   & $2.4\pm 0.1$ & $1.3\pm 0.3$ & $0.17\pm 0.07$
        & \underline{$10.2\pm 1.9$}   & $2.3\pm 0.3$ & $1.3\pm 0.1$ & $0.25\pm 0.05$
        & $9.6\pm 3.4$    & $2.4\pm 0.2$ & \underline{$1.2\pm 0.2$} & $0.28\pm 0.16$ \\
EEGDfus & $\mathbf{8.5\pm 3.1}$   & $2.2\pm 0.1$ & \underline{$0.9\pm 0.1$} & $0.23\pm 0.31$
        & $\mathbf{9.1\pm 3.2}$    & $2.2\pm 0.1$ & $1.4\pm 0.1$ & $0.35\pm 0.23$
        & $\mathbf{8.6\pm 4.2}$    & $2.2\pm 0.1$ & $1.5\pm 0.1$ & $0.16\pm 0.22$ \\
\bottomrule
\end{tabular}
}
\caption{Benchmarking EEG denoising methods across three motion conditions relative to a static intra-subject eyes-closed setting. The results reveal performance degradation under increasing motion and highlight trade-offs between artifact suppression and neural preservation. \textbf{Bold denotes best}, underline denotes second best in each column and motion condition.}
\label{denoise_benchmark}
\end{table*}

\subsubsection{Performance of different approaches}
\label{sec:denoise-performance}
We benchmarked representative methods against a static baseline derived from intra-subject eyes-closed (EC) comparisons: $D_{\text{PSD}} = 4.5 \pm 1.3$, Brain = $0.68 \pm 0.23$, EMG Index = $0.28 \pm 0.06$, and Alpha Consistency = $0.56 \pm 0.45$. Traditional methods (ICA, EMD, ASR) suffer pronounced performance drops as motion becomes more dynamic. For example, ICA's spectral distortion ($D_{\mathrm{PSD}}$) escalates from $16.1$ (body movements) to $50.6$ (navigation). ASR also deteriorates under non-stationary walking and navigation, showing large $D_{\mathrm{PSD}}$ values and unstable EMG suppression. Deep learning models, including GCTNet, EEGDNet, and EEGDfus, consistently achieve low spectral distortion ($D_{\mathrm{PSD}} \approx 9.0$). However, all three exhibit low Brain IC Ratios (e.g., $<0.35$ across conditions), suggesting over-cleaning where meaningful neural components may also be suppressed. These results highlight that linear statistical methods are ill-suited for complex real-world motion, and that multi-metric evaluation is necessary to avoid over-cleaning and ensure neural fidelity.

\subsubsection{Performance under different scenes in navigation tasks}
\begin{figure}
    \centering
    \setlength{\abovecaptionskip}{0.cm}
    \setlength{\belowcaptionskip}{0.cm}
    \includegraphics[width=0.98\linewidth]{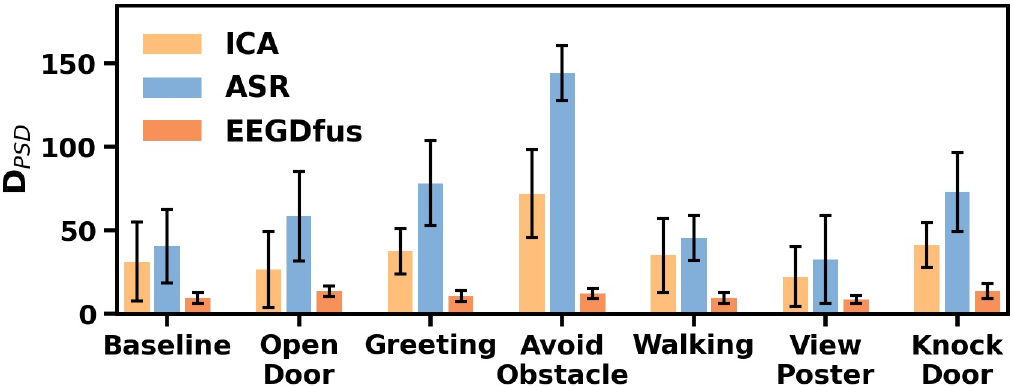}
    \caption{Analysis of noise reduction effects in different scenes of navigation.}
    \label{fig:scene_dpsd}
\end{figure}

We evaluate denoising robustness across navigation scenes (Figure~\ref{fig:scene_dpsd}) by comparing $D_{\mathrm{PSD}}$ against the static baseline. The six scenes are opening a door, greeting an experimenter, avoiding an obstacle, walking along a corridor, reading a wall poster, and knocking on a door. Distortion levels vary systematically with motion type: abrupt or multi-joint actions (e.g., Greeting, Avoid Obstacle) induce the most severe artifacts, while structured scenes (e.g., Open Door, View Poster) show milder deviations. Walking produces moderate errors but still challenges ASR due to gait-induced periodic noise. ICA remains stable yet underperforms; EMD fails consistently under complex motion. EEGDfus shows greater robustness but still degrades as motion becomes less predictable.

\section{CASE STUDY}

We present two case studies enabled by WearBCI. Section~\ref{sec:cs1} investigates whether incorporating auxiliary IMU signals can improve EEG quality under motion, while Section~\ref{sec:cs2} demonstrates that behavioral understanding becomes feasible when multimodal signals are combined.

\subsection{Cross-Modal EEG Signal Enhancement}
\label{sec:cs1}
\begin{figure}
    \centering
    \setlength{\abovecaptionskip}{0.cm}
    \setlength{\belowcaptionskip}{0.cm}
    \includegraphics[width=\linewidth]{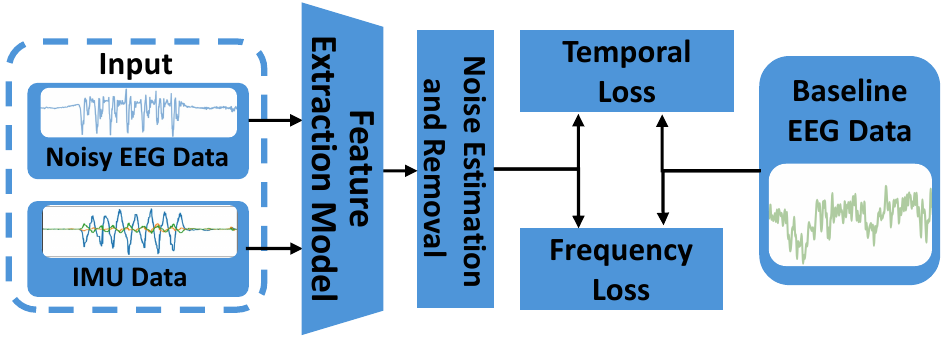}
    \caption{IMU-assisted EEG signal enhancement. }
    \label{fig:enhancement_pipeline}
\end{figure}

To address the limitations of existing denoising methods (Section \ref{sec:denoise-performance}), we investigate whether incorporating motion information can improve robustness. As shown in Figure~\ref{fig:eeg-imu-correlation}, IMU variability correlates with EEG fluctuations, motivating an IMU-assisted enhancement approach.
We adopt a linear regression model that maps multi-site IMU features to motion-induced EEG artifacts in a transparent, model-agnostic manner, as illustrated in Figure~\ref{fig:enhancement_pipeline}. The input consists of acceleration and gyroscope signals from five IMUs placed on the head, wrists, and ankles, and the output is the estimated artifact component for each EEG channel. Training minimizes a combined time-domain MSE and frequency-domain PSD loss, encouraging both waveform fidelity and spectral alignment with static baselines. During inference, predicted artifacts are subtracted from the raw EEG to obtain cleaned signals.

\begin{figure}
    \centering
    \setlength{\abovecaptionskip}{0.cm}
    \setlength{\belowcaptionskip}{0.cm}
    \begin{subfigure}[b]{0.49\linewidth}
        \centering
        \setlength{\abovecaptionskip}{0.cm}
        \setlength{\belowcaptionskip}{0.cm}
        \includegraphics[width=\linewidth]{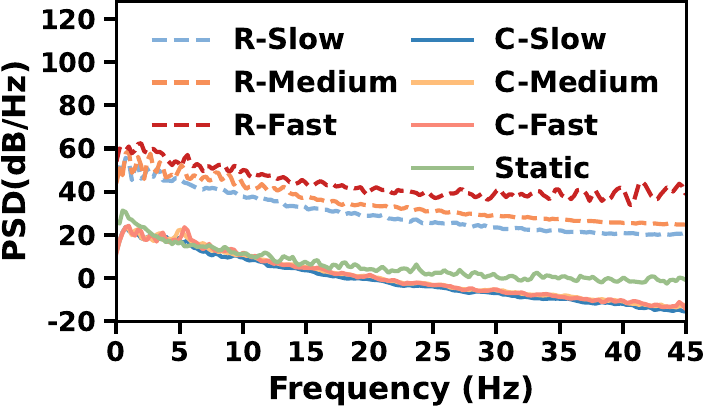}
        \caption{PSD (Frequency domain).}
        \label{fig:psd_enhancement}
    \end{subfigure}
    % \hfill
    \begin{subfigure}[b]{0.49\linewidth}
        \centering
        \setlength{\abovecaptionskip}{0.cm}
        \setlength{\belowcaptionskip}{0.cm}
        \includegraphics[width=\linewidth]{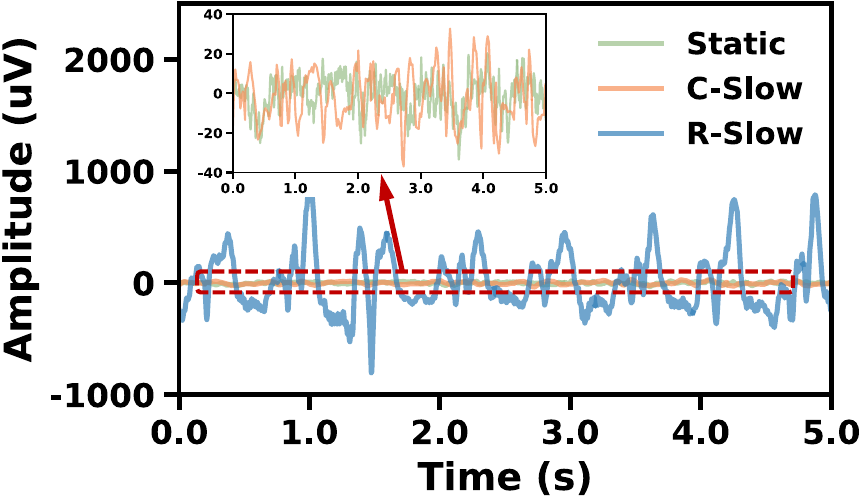}
        \caption{Waveform (Time domain).}
        \label{fig:waveform_enhancement} 
    \end{subfigure}
    \caption{Comparison of PSD and waveform before (R: Raw EEG) and after (C: Clean EEG) IMU-assisted signal enhancement during walking.
    } 
    \label{fig:enhancement}
\end{figure}

Figures~\ref{fig:psd_enhancement} and \ref{fig:waveform_enhancement} show that the cleaned signal better matches the eyes-closed PSD reference with suppression of low-frequency motion bursts, and recovers more stable oscillatory patterns. Quantitatively, our approach reduces spectral distortion ($D_{\mathrm{PSD}} = 10.8 \pm 4.6$) and retains a relatively high proportion of neural components (Brain IC Ratio $= 0.52 \pm 0.21$), outperforming traditional methods in navigation scenes. These results show that even a simple IMU-informed model can enhance EEG quality and motivate future multimodal denoising frameworks for wearable BCIs.

% \vspace{-5pt}

\subsection{Multi-dimension Behavior Understanding}
\label{sec:cs2}

\begin{figure}
    \centering
    \setlength{\abovecaptionskip}{0.cm}
    \setlength{\belowcaptionskip}{0.cm}
    \vspace{-8pt}
    \includegraphics[width=\linewidth]{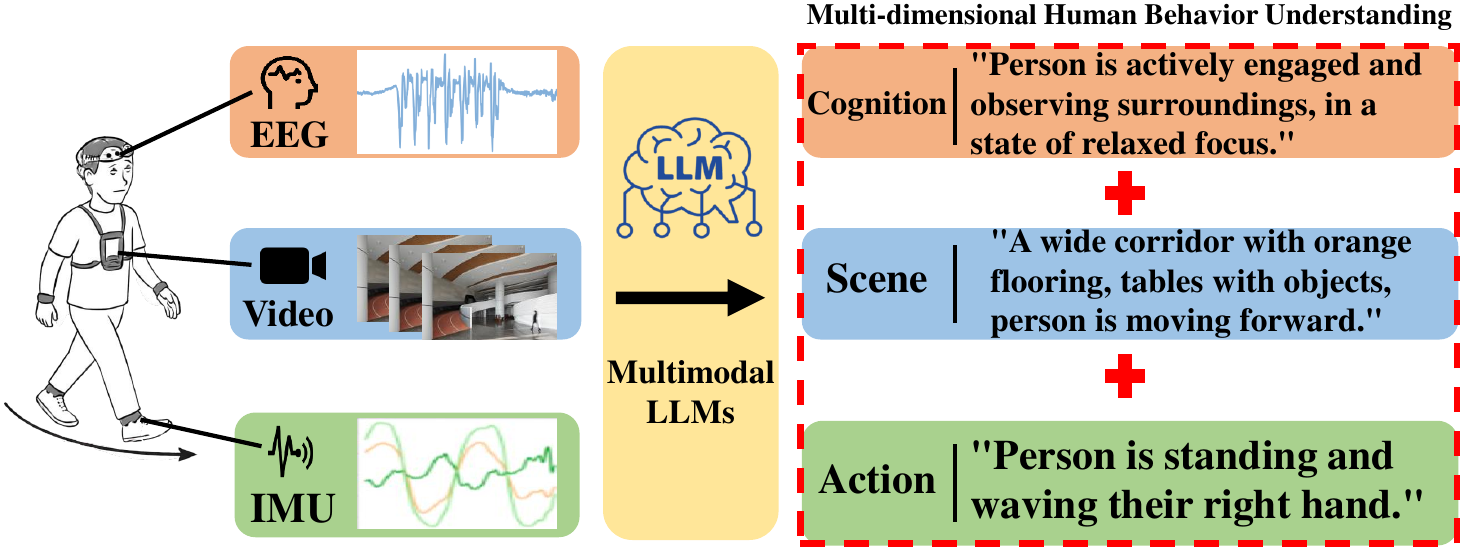}
    \caption{Illustration of multi-dimension behavior understanding through different modalities.}
    \label{fig:egocentric}
\end{figure}

The synchronized multimodal recordings in WearBCI enable richer behavior understanding than any single modality. As illustrated in Figure~\ref{fig:egocentric}, EEG, IMU, and egocentric video each capture distinct dimensions of behavior during naturalistic navigation, where video grounds environmental context, IMU tracks motor dynamics, and EEG reflects cognitive engagement. We annotate the navigation recordings across three semantic categories (Scene, Action, and Cognition) to assess each modality's representational capacity and examine how fusion improves behavioral understanding.

\subsubsection{Data Annotation}
Each navigation recording is segmented based on behavior-driven events (e.g., turning, reading), with annotators marking boundaries based on natural cognitive transitions. Due to individual variability, segmentation differs across participants. We define three semantic categories: \emph{Scene} (the physical environment and context, e.g., ``corridor,'' ``doorway''), \emph{Action} (the motor behavior, e.g., ``opening door,'' ``obstacle avoidance''), and \emph{Cognition} (the inferred mental engagement, e.g., ``visual focus,'' ``spatial planning''), refined into a detailed label set covering edge cases and compound events.

\vspace{-8pt}

\subsubsection{Evaluation}
We use Gemini 2.5 Pro to generate textual descriptions for each modality, and compare them to human annotations using \emph{BERTScore F$_1$} \cite{zhang2019bertscore}. BERTScore is adopted because it provides a context-aware semantic similarity measure in a shared Transformer embedding space, capturing alignment beyond surface-level string matching across heterogeneous sensor inputs. Scores above 0.7 indicate strong alignment, 0.4--0.7 suggest partial overlap, and below 0.4 reflect poor correspondence.
We compute F$_1$ separately for \emph{Scene}, \emph{Action}, \emph{Cognition}, and overall alignment. For EEG and IMU, we use waveform inputs, with EEG additionally including PSD features to enhance decoding.
For modality fusion, we compare two strategies: Concatenate Fusion merges outputs assuming equal importance, while Optimized Fusion applies precision-weighted aggregation followed by cross-modal calibration that reconciles omissions and conflicts.

\subsubsection{Results}
As shown in Table~\ref{tab:multimodal_eval}, modalities differ in representational strength: Video achieves the highest overall BERTScore (0.62), EEG better captures cognitive states (0.45 in Cognition), and IMU best reflects motor actions (0.65 in Action). Fusion further improves—Concatenate Fusion reaches 0.73, and Optimized Fusion reaches 0.86 by reconciling cross-modal conflicts. These results demonstrate that EEG, IMU, and video capture distinct but synergistic aspects of behavior, and that multimodal integration compensates for the limited spatial coverage of wearable EEG.

\begin{table}
\centering
        \setlength{\abovecaptionskip}{0.cm}
        \setlength{\belowcaptionskip}{0.cm}
\small
\begin{tabular}{lcccc}
\toprule
\textbf{Modality} & \textbf{Cognition} & \textbf{Scene} & \textbf{Action} & \textbf{Overall} \\
\midrule
EEG-only          & 0.451  & 0.271          & 0.573      & 0.386 \\
IMU-only          &    0.168        & 0.143          & 0.649     & 0.330 \\
Video-only        &    0.276        & 0.864 &    0.652           & 0.622 \\
Concatenate Fusion&    \textbf{0.477}        & 0.885          &    0.229           & 0.727 \\
Optimized Fusion  &    0.413        & \textbf{0.907}          &    \textbf{0.729}           & \textbf{0.859} \\
\bottomrule
\end{tabular}
\caption{Quantitative evaluation of semantic alignment (BERTScore F$_1$) across modalities and fusion strategies.}
\label{tab:multimodal_eval}
\end{table}

\section{Discussion}

\textbf{Key insights and lessons learned.} First, motion severely degrades EEG signal quality of wearable BCI systems, with spectral distortion increasing over 10-fold from static to navigation conditions. Second, existing denoising methods struggle under realistic motion conditions: traditional methods such as ASR and EMD experience catastrophic failure as motion complexity increases, as stationarity assumptions break down; deep learning models reduce spectral distortion more effectively but suppress neural components excessively. Third, the synchronized multimodal recordings in WearBCI enable signal enhancement and behavioral understanding beyond single-modality EEG, as shown in our case studies.

\noindent \textbf{Future directions and enabled applications.} First, our case study shows that multimodal sensing can infer internal states like attention during natural behavior, enabling more personalized interventions for populations such as those with Alzheimer's disease. Moreover, as discussed in Section~\ref{sec:region}, different actions contaminate distinct brain regions—for example, knocking affects frontal areas while walking introduces global interference—suggesting the value of region-aware models. While the current dataset provides systematic coverage of representative motion scenarios under controlled indoor conditions, future work should extend to unconstrained outdoor settings and more diverse participant populations to further improve ecological validity.

\section{Conclusion}
In this work, we introduce WearBCI, a multimodal dataset containing synchronized EEG, IMU, and egocentric video from 36 participants across movement scenarios of increasing complexity.
Through benchmark experiments and case studies, we examine how motion impacts EEG quality, evaluate current EEG enhancement methods, and show that multimodal signals enable both cognitive decoding and signal enhancement. WearBCI is designed as a benchmarking resource for the community to develop and evaluate approaches for enhancing EEG signals collected by wearable BCI systems, and supports the development of more robust wearable BCI systems for everyday use.d

\section*{ACKNOWLEDGMENTS}
This work is supported by the Research Grants Council (RGC) of Hong Kong, China, under grant ECS 26200825, the HKUST-HKUST(GZ) Cross-campus Research Collaboration ``1+1+1'' Joint Funding Program under G\_2025\_052, and is partly funded by the HKUST Institute for Emerging Market Studies with support from EY, under grant IEMS25EG01.

\clearpage
\FloatBarrier
\bibliographystyle{ACM-Reference-Format}
\bibliography{reference}

\end{document}